\documentclass[10pt, notitlepage, twoside, a4paper]{article}
\usepackage{a4wide, amssymb, epsfig, graphicx, fancyhdr}

\newcommand{\Section}[1]{\section{#1}\vspace{-8pt}}

\title{Hard Pion Chiral Perturbation Theory: What is it\\and
is it relevant for $\eta^\prime$ decays?}

\author{JOHAN BIJNENS\footnote{Department of Astronomy and Theoretical Physics, Lund University, S\"olvegatan 14A, SE 223 62 Lund, Sweden, bijnens@thep.lu.se}
}

\date{}

\begin{document}
\rhead{}
\lhead{}
\pagestyle{fancy}
\renewcommand{\headrulewidth}{0.2pt}
\renewcommand{\footrulewidth}{0.2pt}
\begin{titlepage}
\large
\begin{flushright}
LU TP 11-39\\
October 2011
\end{flushright}
\vfill
\begin{center}
{\Large\bf Hard Pion Chiral Perturbation Theory: What is it\\[4mm]
and is it relevant for $\eta^\prime$ decays?\footnote{Talk presented at the PrimeNet meeting,
Meson Physics in Low-Energy QCD, September 26 - 28, 2011,
Forschungszentrum J\"ulich, Germany.}}\\[2cm]
{\bf Johan Bijnens}\\[0.5cm]
Department of Astronomy and Theoretical Physics,\\[2mm]Lund University,
S\"olvegatan 14A, SE 223 62 Lund, Sweden
\end{center}
\vfill

\begin{abstract}
In this talk I give a short introduction to hard pion Chiral Perturbation Theory
and an overview of the available applications
$K\to\pi\pi$, $B,D\to D,\pi,K,\eta$ semileptonic decays and
$\chi_{c0,2}\to\pi\pi,KK$. It is pointed out that the reults
for the semileptonic decays obey the LEET relation between $f_+$ and $f_-$.
\end{abstract}

\vfill

\end{titlepage}

\maketitle

\Section{Introduction}

In this talk I will try to convince you that we can give predictions from
chiral symmetry also for cases where not all pions are soft.
This is something I called hard pion Chiral Perturbation Theory (HPChPT)
and there have been a few recent papers using this
\cite{FS,BC,BJ1,BJ2,BJ3}.

I will first give a short introduction to effective field theory (EFT)
and remind
you of the underlying principles of Chiral Perturbation Theory (ChPT).
I will remind you of the fact that in ChPT with baryons and other
heavy particles a power-counting has been achieved by consistently
absorbing the heavy mass dependence into the low-energy-constants (LECs).

The arguments will then be generalized to the case of processes with high energy
or hard pions. The arguments also apply to cases where we can
treat the strange quark mass as small as well.

After that I will show applications to $K\to\pi\pi$, to semileptonic decays
of pseudo-scalar mesons or to more general vector form-factors and to
charmonium decays to two pseudo-scalars.

Unfortunately, there seem to be no $\eta^\prime$ decays where the present method
are applicable.

\Section{Effective field theory and ChPT}

The underlying idea of EFT is a general theme in science,
restrict yourself to the relevant degrees of freedom.
So, in cases where there exists an energy or mass gap we keep only
the lower degrees of freedom. Lorentz-invariance
and quantum mechanics imply that we are restricted to a field theory
but we should build the most general one with our chosen degrees of freedom.
We have no predictability left since the most general
Lagrangian will have an infinite number of parameters. This can be cured
if we find an ordering principle, power-counting, for the
importance of terms.

ChPT is ``exploring the consequences of the chiral symmetry of QCD and its
spontaneous breaking using effective field theory techniques'' and
was introduced as an EFT in \cite{Weinberg0,GL1}.
The degrees of freedom are the Goldstone bosons from the
spontaneous breakdown of the global chiral $SU(n)_L\times SU(n)_R$
to the diagonal vector subgroup $SU(n)_V$. That the Goldstone boson
interactions vanish at zero momentum  allows to construct a
consistent power-counting
\cite{Weinberg0}.

The basic form of ChPT has since been extended to baryons,
mesons and baryons containing a heavy quark, vector mesons,
structure functions and related quantities as well as beyond the pure
strong interaction by including weak and electromagnetic internal interactions.
Many models of alternative Higgs sectors also use the same technology.

\Section{Power-counting and one large scale}

\begin{figure}
\centerline{\includegraphics[width=0.7\textwidth]{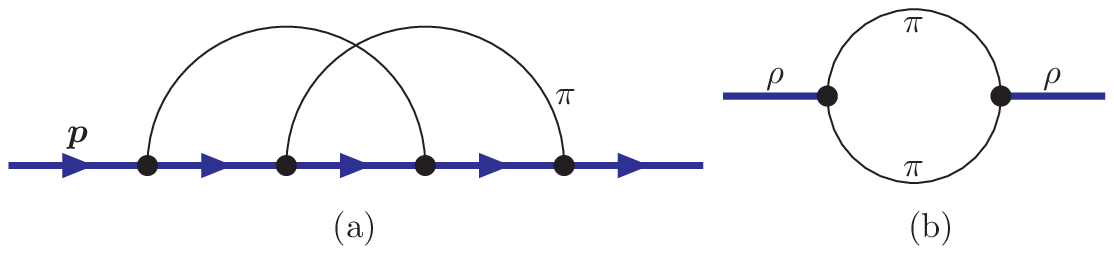}}
\caption{\label{figbaryonChPT}
(a) A typical baryon ChPT diagram with the baryon going through
the entire diagram. (b) An example of a diagram in vector meson ChPT
with no continuous vector meson line.}
\end{figure}

In purely mesonic ChPT the power-counting is essentially
dimensional counting and this works since all the lines in all diagrams have
``small'' momenta. Already when discussing baryons, this lead to problems
because now there is a large scale, the baryon mass.
However, by setting the baryon momentum $p_B= M_B v+k$ with $v$
the baryon four-velocity, a consistent
power-counting can be achieved. This works ``obviously'' since the
heavy line goes through the entire diagram and all momenta apart
from $M_B v$  are soft as indicated by the thick line in
Fig.~\ref{figbaryonChPT}(a).
The same arguments apply to ChPT for mesons containing a heavy quark.
It works because the ``soft'' stuff can be expanded
in ``soft$/M_B$'' and the remaining $M_B$ dependence can be absorbed in
LECs.
For vector meson ChPT there is a problem since they can decay.
A typical diagram is shown in Fig.~\ref{figbaryonChPT}(b). However it was
argued that the non-analytic dependence on the light quark mass could still
be obtained, see the discussions in \cite{BGT}. Again, the underlying idea
is that the large $M_V$ allows to expand in ``soft/$M_V$''
and the remaining $M_V$ dependence is absorbed in the LECs.


\Section{Several large scales or HPChPT}

In \cite{FS} the authors applied heavy Kaon ChPT to $K_{\ell3}$
at the endpoint, i.e. the pion is soft, this works as in usual ChPT.
They also applied it to the region
for small $q^2$ where the pion has a large momentum and gave arguments
based on partial integrations why this would give a correct chiral
logarithm. The argument was generalized in
\cite{BC,BJ1,BJ2,BJ3}. The underlying idea is similar to the
previous
section. The ``heavy/fast/hard'' dependence on the soft stuff can always
be expanded and the remaining dependence goes into the LECs.
That this might be possible follows also from current algebra. Non-analyticities
in the light masses come from the soft lines and soft pion couplings are
restricted by current algebra via
$
\lim_{q\to 0}\langle\pi^k(q)\alpha|O|\beta\rangle
 = -\frac{i}{F_\pi}\langle\alpha|\left[Q_5^k,O\right]|\beta\rangle\,.
$
Nothing prevents hard pions to be in the states $\alpha$ or $\beta$,
so by heavily using current algebra one can get
the light quark mass non-analytic dependence

A field theoretic argument is:
(1) Take a diagram with a given external and
internal momentum configuration. (2)
Identify the soft lines and cut them. (3)
The resulting part is analytic in the soft stuff,
so it can be described by an effective Lagrangian with
coupling constants dependent on the external given momenta
(Weinberg's folklore theorem \cite{Weinberg0}).
(4) The non-analytic dependence on the soft stuff is reproduced by loops
in the latter Lagrangian.
\begin{figure}
\centerline{\includegraphics[width=0.7\textwidth]{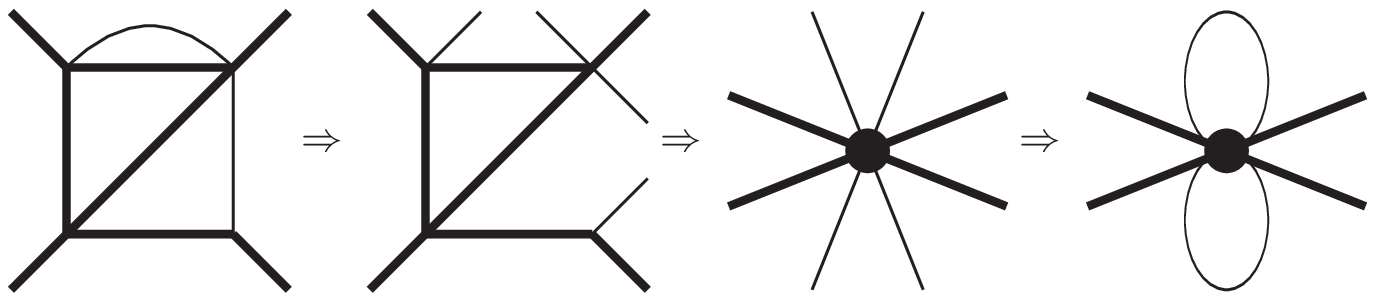}}
\caption{\label{figcuthardlines} The process of cutting the soft lines
and reproducing the non-analytic dependence by the diagram on the right.
The hard lines are thick, soft lines are shown thin.}
\end{figure}
The process is depicted in Fig.~\ref{figcuthardlines}.

The remaining problem is that we have no power-counting. The
Lagrangian that reproduces the non-analyticities is fully
general. In the HPChPT papers it was shown that for the
processes at hand, all higher order terms can be reduced
to those with the fewest derivatives thus allowing the light quark mass
chiral logarithm to be predicted. The underlying arguments were tested
by comparing to a two-loop calculation\cite{BJ2}
and by explicitly keeping some higher order terms \cite{BC,BJ1,BJ2,BJ3}.

\Section{Applications}

We have applied the method to $K\to\pi\pi$ decays \cite{BC}
where we treat the Kaon as heavy and look for the dependence on
the pion mass $M^2$. The result is, up to linear in $M^2$ and higher order:
$$
A_0^{NLO}/A_0^{LO} =1+({3}/{8})\hat A\,,
\quad\!\!
A_2^{NLO}/A_2^{LO}=1+({15}/{8})\hat A\,,
\quad\!\!
\hat A = -{M^2}\ln({M^2}/{\mu^2})/({16\pi^2 F^2}).
$$

The scalar and vector form-factors of the pion are known to two-loops in
ChPT \cite{BCT}. HPChPT predicts at large $t$ \cite{BJ2} with
$F_V(t,0)$ and $F_S(t,0)$, the form-factors at large $t$ in the chiral limit,
completely free:

$$
F_V(t,M^2) = F_V(t,0)\,
(1+\hat A)\,,\quad
F_S(t,M^2) = F_S(t,0)\,
(1+({5}/{2})\hat A)\,.
$$
The full two-loop result expanded for large $t$ should have this form
and it does with for e.g. $F_V$
$$
F_V(t,0) = 1+ ({t}/({16\pi^2F^2}))
\left({5}/{18}-16\pi^2 l_6^r+{i\pi}/{6}-({1}/{6})\ln({t}/{\mu^2})
 \right)\,.
$$

\begin{figure}
\begin{minipage}{0.49\textwidth}
\includegraphics[angle=270, width=0.99\textwidth]{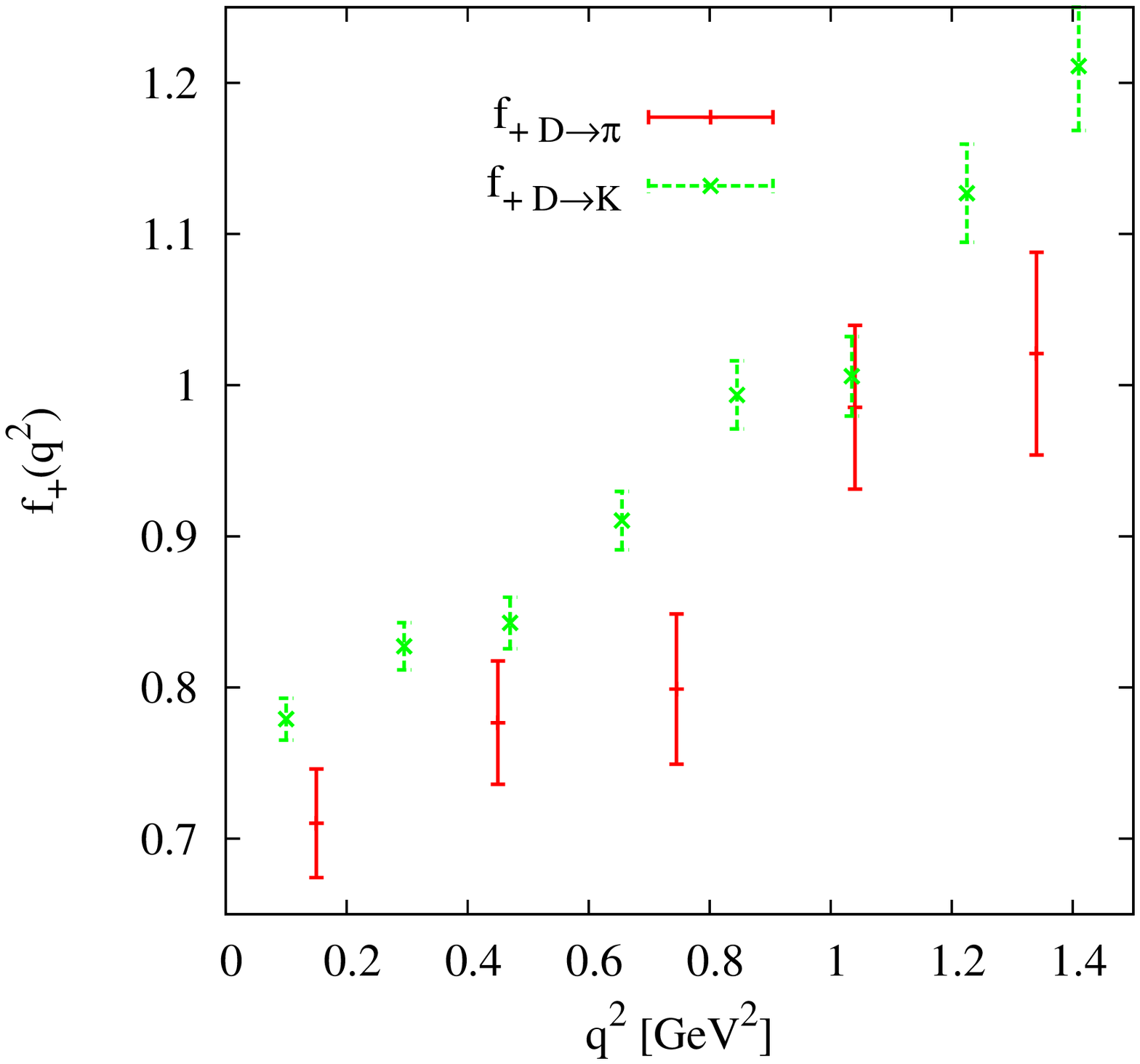}
\end{minipage}
\begin{minipage}{0.49\textwidth}
\includegraphics[angle=270, width=0.99\textwidth]{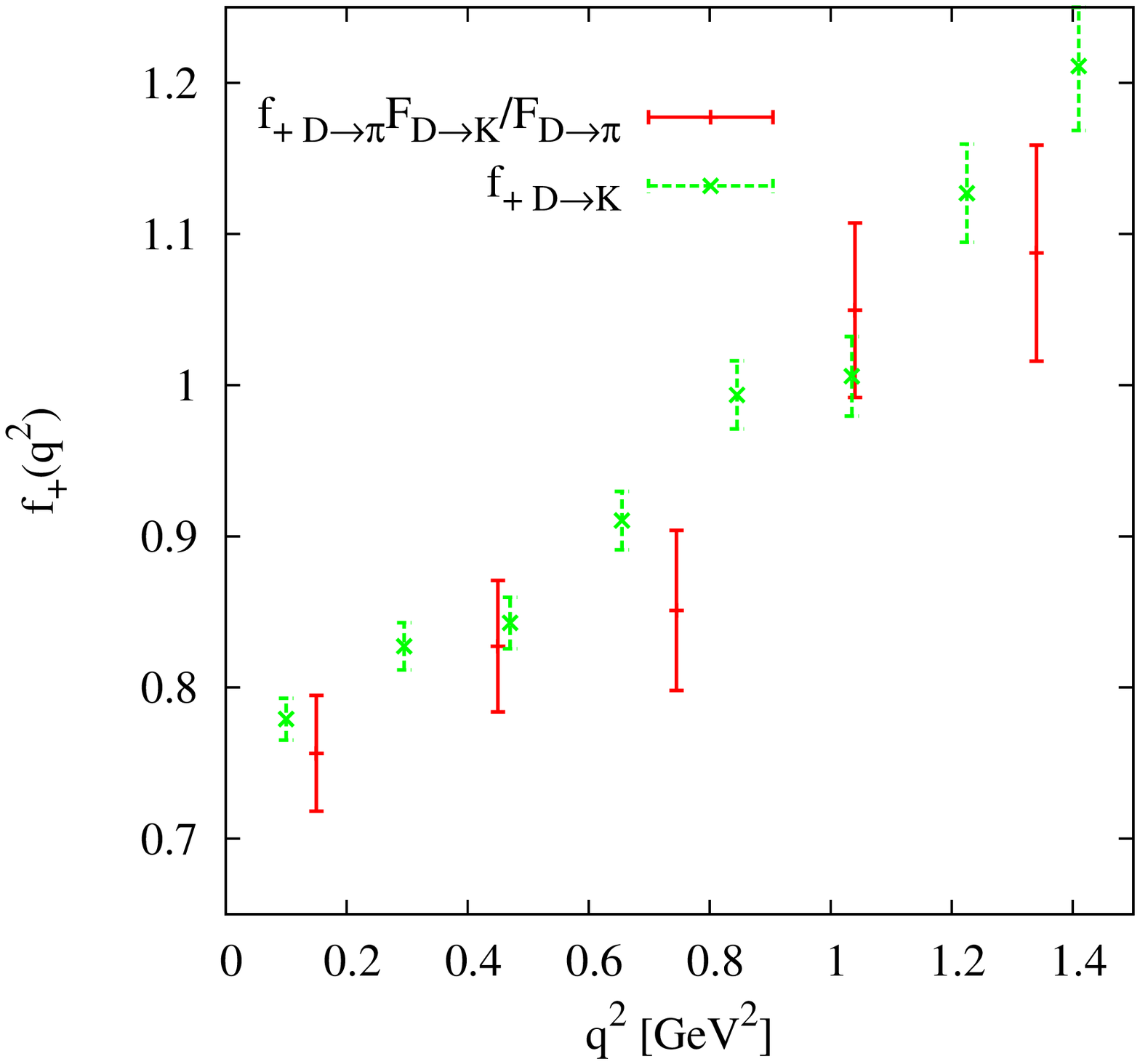}
\end{minipage}
\caption{\label{figfvpCLEO} The CLEO data on $D\to\pi$ and $D\to K$.
Form-factors as measured to the the left and corrected with the chiral
logarithms to the right. Note the improved agreement.}
\end{figure}

The first application was semileptonic form-factors in $K_{\ell3}$.
We extended this to $B,D\to D,\pi,K,\eta$-decays in \cite{BJ1,BJ2}.
This allowed to test our results experimentally. The form factor $f_+(t)$
as measured by CLEO \cite{CLEO} in $D\to\pi$ and $D\to K$ decays are different
by about the amount expected from the chiral logarithms as shown in
Fig.~\ref{figfvpCLEO}.
One puzzling observation \cite{BJ2} was that in the limit of a
hard pseudo-scalar
in the final state the correction was always the same for $f_+$ and $f_-$.
This is in fact due to the LEET relation \cite{LEET} which shows that there
is only one form-factor in this limit and it is nice to see that our
calculation respects this without it being used as input.

A last application was to $\chi_{c0,2}\to\pi,KK,\eta\eta$ \cite{BJ3}.
Here it was found that there were no chiral logarithms to the order considered.
Comparing with the known experimental results indeed shows $SU(3)$ breaking
to be somewhat smaller than in e.g. $F_K/F_\pi$. Details can be found
in \cite{BJ3}.

\section*{Acknowledgments}\vspace{-8pt}
This work is supported in part by the EU, 
HadronPhysics2 Grant Agreement n. 227431,
and the Swedish Research Council, grants 621-2008-4074 and 621-2010-3326.

\end{document}